\documentclass[12pt,twocolumn]{iopart}
	
	\usepackage{amssymb}
	\usepackage[]{graphicx}

	\newcommand{\beq}{\begin{equation}}
	\newcommand{\eeq}{\end{equation}}
	\newcommand{\bea}{\begin{eqnarray}}
	\newcommand{\eea}{\end{eqnarray}}

\begin{document}

	\title{Vortex kinks in superconducting films with periodically modulated thickness}

	\author{Jorge I. Facio$^{1,2}$, Anabella Abate$^{1}$, J. Guimpel$^{1,2,3}$, Pablo S. Cornaglia$^{1,2,3}$}
	\address{$^1$Instituto Balseiro, Comisi\'on Nacional de Energ\'{\i}a At\'omica and Universidad Nacional de Cuyo, Centro At\'omico Bariloche, 8400 Bariloche, Argentina}
	\address{$^2$Consejo Nacional de Investigaciones Cient\'{\i}ficas y T\'ecnicas (CONICET), Av. Rivadavia 1917, Buenos Aires, Argentina}
	\address{$^3$Centro At{\'o}mico Bariloche, Comisi\'on Nacional de Energ\'{\i}a At\'omica, 8400 Bariloche, Argentina}

	\begin{abstract}
	We report magnetoresistance measurements through Nb films having a periodic thickness modulation. The cylinder shaped large-thickness regions of the sample, which form a square lattice, act as repulsive centers for the superconducting vortices. For low driving currents along one of the axes of the square lattice, the resistivity $\rho$ increases monotonously with increasing magnetic field $B$ and the $\rho$--$B$ characteristics are approximately piecewise linear. The linear $\rho$ vs $B$ segments change their slope at matching fields where the number of vortices is an integer or a half integer times the number of protruding cylinders in the sample.
	Numerical simulations allow us to associate the different segments of linear magnetoresistance to different vortex-flow regimes, some of which are dominated by the propagation of discommensurations (kinks).
	\end{abstract}

	\maketitle

	\section{Introduction}
	The dynamics of driven superconducting vortices has been the subject of intense theoretical and experimental research during the last decade. This has been mainly fueled by the technological interest in reducing the dissipation associated with the vortex motion and by the possibility of using vortex matter as a model system with tunable parameters to study static and dynamic phases and their transitions. In vortex systems it is possible to control the driving force by applying an electric current across the sample, to set the density through external magnetic fields, and to tailor the potential landscape using lithographic techniques. 

	The fact that the density of vortices in a sample can be tuned applying an external magnetic field, has allowed a number of studies that analyze commensurability effects between the vortex lattice and an underlying periodic potential. The typical matching signature is the appearance of minima in the resistivity (or maxima in the critical current) for magnetic fields such that 
	the number of vortices $n_V$ is a multiple (or a simple fraction like 1/4, 1/2 and 3/4) of the number of minima $n_P$ in potential landscape.
	For this purpose, films having different types of artificial pinning centers like holes~ 
\cite{fiory:73,Baert_holes} magnetic 
\cite{PhysRevLett.79.1929,MorganDipoles} or non-magnetic particles 
\cite{PhysRevB.61.6958} in different geometries 
\cite{HaradaScience,
MatsudaScience,
PhysRevB.58.8232,
VelezRectang,
MetlushkoSquare,
VanBael200012,
Martin2000,
Crabtree2002,
PhysRevB.65.104518,
FasanoSquare,
Chialvo2005112,
FasanoPNAS,
PhysRevB.72.014507,
silhanek:152507,
0953-8984-21-7-075705,
springerlink:10.1007/s10909-011-0450-1},
have been studied both experimentally and theoretically 
\cite{
PhysRevLett.78.2648,
PhysRevB.58.6534,
Nori1999,
PhysRevLett.83.3061,
ReichThermo,
PhysRevB.64.104505,
LagunaTriang,
MoshchalkovSquare,
CornagliaSquare}.

	Driven vortices in periodic potentials present a rich variety of phenomena~\cite{Velez20082547}, including transitions between different vortex flow regimes. A prominent example is the rich phase diagram predicted by Reichhardt {\it et al.}~\cite{PhysRevLett.78.2648,PhysRevB.58.6534} for a square pinning potential by means of numerical simulations, which was later confirmed experimentally (see reference \cite{PhysRevB.80.140514}).

	A common feature in the general problem of interacting particles in a periodic potential is the formation of discommensurations in the particle lattice \cite{Floria_Mazo_1996}.
	These appear as a result of the competition between inter-particle interactions, which favor a specific crystal structure, and the external potential, which in general, favors another. In vortex systems these discommensurations have been predicted in numerical simulations \cite{PhysRevB.58.6534,PhysRevB.83.184514,Pogosov2010} and also visualized experimentally \cite{0295-5075-54-5-682}, but their dynamics and the experimental consequences of their presence have been less explored.

	In this work we analyze the dynamics of driven vortices in a square pinning potential as a function of the vortex density and driving force. To that aim, we generate a periodic thickness modulation in Nb films and perform magnetoresistance measurements in the superconducting state. The potential energy of a vortex increases with the sample thickness due to the extra energy associated with the vortex length~\cite{PhysRevLett.32.218,PhysRevB.77.024526}. This allows to control the intensity and the topology of the pinning potential energy using lithographic techniques.

	For a film with a square lattice of protruding cylinders we observe clear commensurability signatures. The resistivity increases linearly with increasing magnetic field, but instead of the usual minima at the matching fields we observe changes in the slope of the $\rho$ vs $B$ curve. To interpret the results we perform molecular dynamics simulations, and show that the observed behavior can be ascribed to different regimes of vortex propagation. A low field regime that can be understood in terms of single vortex physics, and higher field regimes dominated by the presence of moving discommensurations (kinks).  

	The rest of the paper is organized as follows: 
	In section \ref{sec:exp} we describe the experimental setup. 
	In section \ref{sec:mr} we present magnetoresistance measurements illustrating the commensuration effects.
	In section \ref{sec:methods} we present the model and methods to describe the driven vortex dynamics. 
	In section \ref{sec:simulations} we present the results of numerical simulations that reproduce qualitatively the experimental observations. 
	In section \ref{sec:toymodel} we present a simplified one-dimensional (1D) model that contains the minimum ingredients to understand the kink dynamics. 
	Finally, in section \ref{sec:concl} we present our concluding remarks.

	\section{Experiment}\label{sec:exp}

\begin{figure}[t]
\begin{center}
	 \includegraphics[width=8 cm,angle=0,keepaspectratio=true]{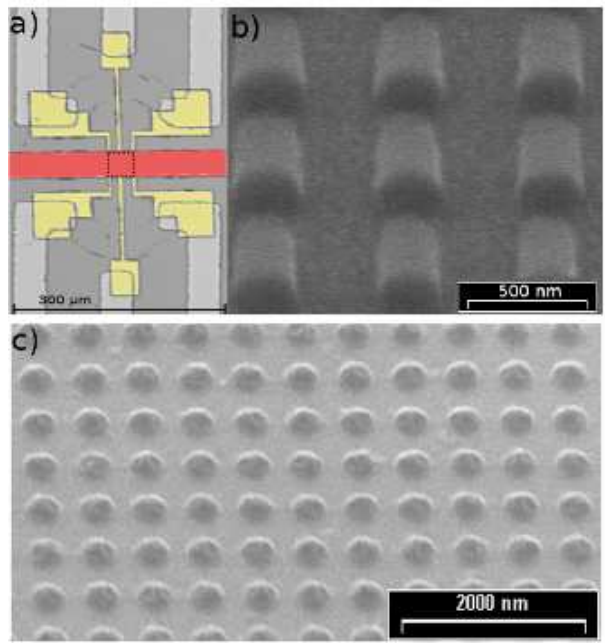}
	\caption{a) Scanning electron microscopy image of the measurement circuit. The current path (colored in red) is coupled to contacts for voltage measurement (colored in yellow). The box of dotted style lines in the current path indicates the region where thickness modulation is performed. 
b) $45^\circ$ tilted scanning electron microscopy image of the square lattice of resist cylinders before performing the RIE. The dark regions correspond to the bases of the cylinders.
c) Image of a sample with Nb cylinders of diameter $d =\, 350 nm$ and lattice parameter $a =\, 570 nm$.}
\label{fig:fabrication}
\end{center}
\end{figure}

	We fabricated a circuit (see figure \ref{fig:fabrication}) consisting of an electrical current path $60 \mu m$ wide coupled to electrodes for longitudinal voltage measurement~\cite{Chialvo2005112}. 
	This structure is fabricated in two stages. First, the desired pattern 
	for the extension electrodes is defined using optical lithography on a AZ-9260 resist layer. Nb is sputtered on top 
	and a lift-off process is performed. Second, using the Nb deposited as an alignment mark, the pattern for the current path and 
	the first part of the contacts is written with e-beam lithography on a bilayer of MMA / PMMA resists. Then again, a  
	Nb film is sputtered on top and a lift-off process is performed.

The periodic modulation of the thickness is generated on $60 \mu m$ long regions located on the current path. Using the negative photo-resist 
ma-N 2403 a square lattice of resist cylinders [see figure \ref{fig:fabrication}(b)] is fabricated by e-beam lithography. When the sample is exposed to reactive ion etching (RIE), the cylinders protect the covered portion of the film, generating the desired thickness modulation. 
During the RIE process we work with a mixture of \emph{SF}$_{6} $ and \emph {Ar} at a pressure of $1.33Pa$, and an acceleration voltage for the ions of approximately $110 V$. With these parameters we obtain a Nb etching speed of $80 nm/min$. After the RIE process, 
the sample is immersed four hours in acetone in order to remove the resist. Figure \ref{fig:fabrication}(c) shows a 
scanning electron microscopy image of one sample with cylinders of diameter $d = 350 nm$ and lattice parameter $a = 570 nm$. As usual, $a$ determines the first matching field $B_\phi = \phi_0 / a^2$, where  $\phi_0$ is the flux quantum.
At $B=B_\phi$ there are as many vortices as unit cells in the cylinder lattice (i.e. $n_V=n_P$).

In what follows we present magnetoresistance measurements for two samples: sample I with $B_\phi=62\,G$ and $T_c=(7.55\pm0.02)K$, and sample II in which $B_\phi=48\,G$ and $T_c=(7.68\pm0.02)K$. The film thickness is $150\,nm$ and $110\,nm$ for samples I and II respectively. In both samples the nominal cylinder's height with respect to the film is $h_{cyl} = 40\,nm$.

\section{Magnetoresistance measurements} \label{sec:mr}

The magnetoresistance measurements were carried out in a four probe geometry within a pulse tube refrigerator with $2.3K$ base temperature. The voltage was measured at constant temperature $T$ and current $J$ along the current path (see figure \ref{fig:fabrication}) using a nanovoltmeter. The magnetic field was increased from zero up to a maximum value of $B_{max}\sim 3B_\phi$ and then decreased to zero again. No hysteresis effects were observed.
\begin{figure}[t]
\begin{center}
 \includegraphics[width =8.5cm]{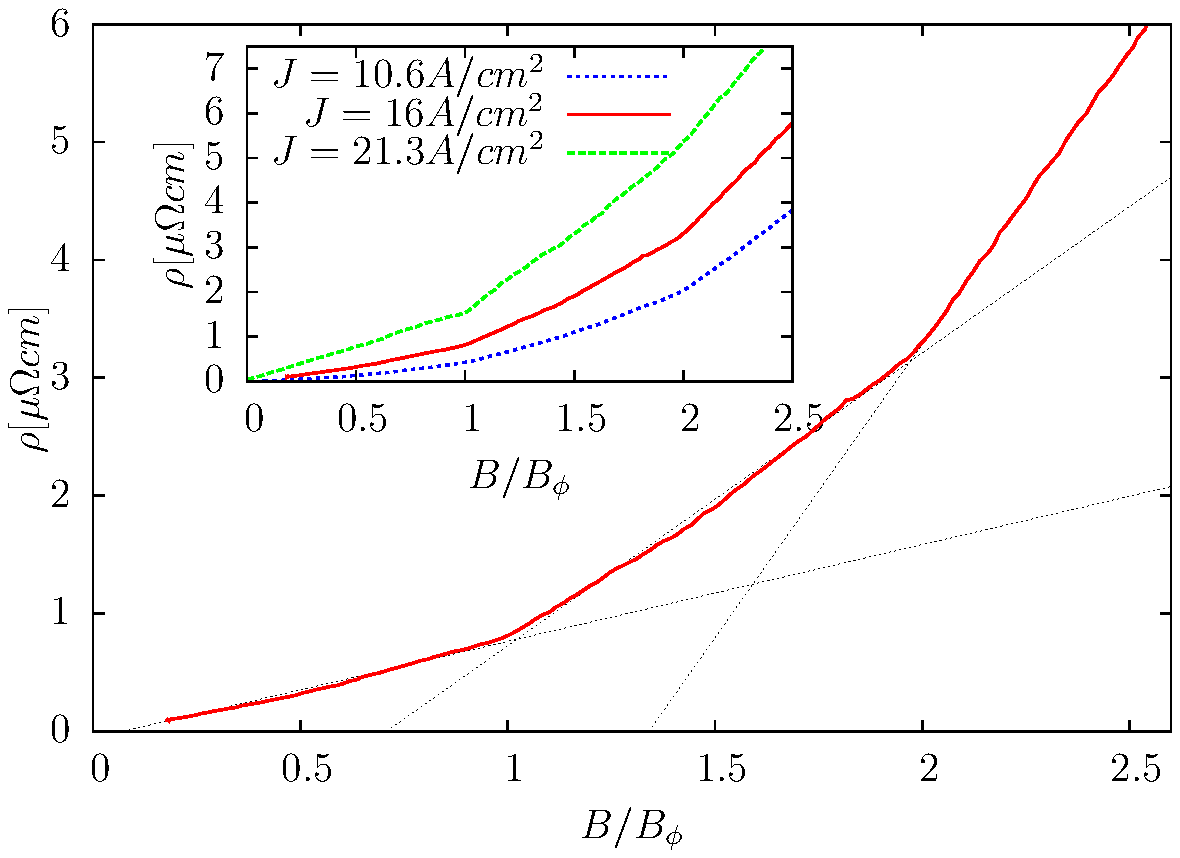}
\caption{Resistivity vs magnetic field in a Nb film with a square lattice of protruding cylinders at $T=7.52K$ ($T/T_c=0.996$) (sample I). The lattice parameter is $a=570nm$ ($B_\phi=62\, G$) and the diameter and height of the cylinders is $d=350nm$ and $h_{cyl}=40nm$, respectively. The resistivity presents an approximately piecewise linear behavior with slope changes at the first two matching fields. As the current driving the vortices is increased (see inset) the change in the slope at the second matching field $B=2 B_\phi$ becomes less pronounced and eventually disappears.}
\label{fig:SlopesJ} 
\end{center}
\end{figure} 

\begin{figure}[t]
\begin{center}
 \includegraphics[width =8.5cm]{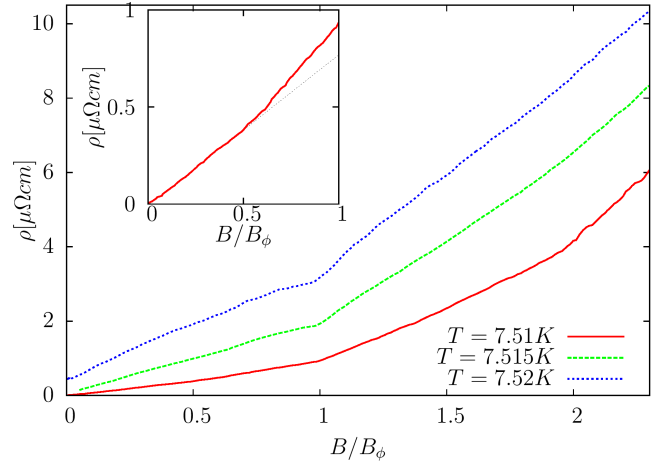}
\caption{Resistivity vs magnetic field in a Nb film with a square lattice of protruding cylinders for $J=26.6 A/cm^2$ and different values of the temperature (sample I). Inset: low field detail of the magnetoresistance at T=7.51K, presenting a commensurability effect at $B=B_\phi/2$. 
}
\label{fig:SlopesT} 
\end{center}
\end{figure} 

In figures \ref{fig:SlopesJ} and \ref{fig:SlopesT} we present magnetoresistance measurements for sample I. 
We focus our analysis on the data obtained within the range of temperatures $7.5K<T<7.52K<T_c$, where $T_c\simeq 7.55K$ is the superconducting critical temperature of the sample, and the range of current densities $10 A/cm^2<J<27 A/cm^2$. 
For lower temperatures or currents the commensurability effects are weak, presumably due to the dominance of the intrinsic pinning mechanisms over the periodic potential \cite{PhysRevB.61.R9249,PhysRevB.66.212507,0953-2048-24-6-065008}. For temperatures or currents above the selected intervals, we observed a finite dissipation at zero external magnetic field which indicates the presence of vortices induced by the external current or other dissipation mechanisms.

At the lowest temperatures and magnetic fields considered, we observe commensurability effects in the magnetoresistance at the matching fields $B=B_\phi$, $B_\phi/2$, $3B_\phi/2$ and $2B_\phi$  (see figures \ref{fig:SlopesJ} and \ref{fig:SlopesT}). Instead of the usual minima in the $\rho$ vs. $B$ curves (see e.g. reference \cite{PhysRevB.65.104518}), we observe a monotonously increasing piecewise linear behavior where the commensurability manifests itself as a series of increases in the slope of the curves at the matching fields. This behavior was observed in different samples obtained using films of various thicknesses and samples produced with different cylinder lattice parameters on a single film, and is presumably associated to the regularity and the low strength of the pinning potential generated by the thickness modulation and the parameter regime of the experiments.

The linear behavior of $\rho(B)$ at low fields is easily understood under the assumption that the periodic pinning potential due to the thickness modulation of the films has a negligible disorder both in the position of the repulsive centers and their amplitude. For low enough magnetic fields $B\to 0$, the average inter-vortex distance $\langle d\rangle$ is large compared to the penetration depth $\lambda$, and the typical vortex-vortex interaction is exponentially small $\propto e^{- d/\lambda}$ ~\cite{BlatterRevModPhys}. In this regime, the contribution from each vortex to the resistivity is magnetic field independent and proportional to its average velocity $\langle v\rangle$ in a direction perpendicular to the external current. If the distribution of $\langle v\rangle$ is sharply peaked at a finite value $\langle v\rangle_{sv}$, the resistivity is simply proportional to the number of vortices ($n_V = A\, B/\phi_0$, where $A$ is the area of the film) 
multiplied by $\langle v\rangle_{sv}$ and a linear increase in the resistivity with increasing magnetic field is obtained: $\rho \propto B\, \langle v\rangle_{sv}$. 
In the experiments, the approximately linear behavior extends up to the field $B=B_\phi$ (or $B=B_\phi/2$ depending on the temperature and the external current), where a change in the slope is observed giving rise to a new linear behavior for higher fields. 

As we will see below through numerical simulations, the persistence of the linear behavior for fields where the vortex-vortex interaction is no longer negligible and the changes in the slope of the magnetoresistance at the matching fields can be understood in terms of vortices moving along channels between pairs of cylinder rows. 
At the matching field $B=B_\phi$, the vortices in a given channel are evenly spaced and move with an average velocity $\sim\langle v\rangle_{sv}$. Adding a vortex to the channel generates a defect (kink) in the vortex lattice that travels faster along the channel than the individual vortices and produces an increase in the dissipation. Quite generally, adding vortices to a channel in a commensurate situation creates kinks that travel faster than the individual vortices. The additional dissipation produced by the moving defects is at the origin of the increase in the slope of the magnetoresistance, which is roughly proportional to the velocity of the kinks $v_k$. 
At the higher matching fields, the vortices are more densely packed and the average inter-vortex interaction is greater. As a consequence, the kinks generated over the following matching fields have larger velocities, leading to the observed increases in the slope of the magnetoresistance.

\begin{figure}[t]
\begin{center}
 \includegraphics[width =8.5cm]{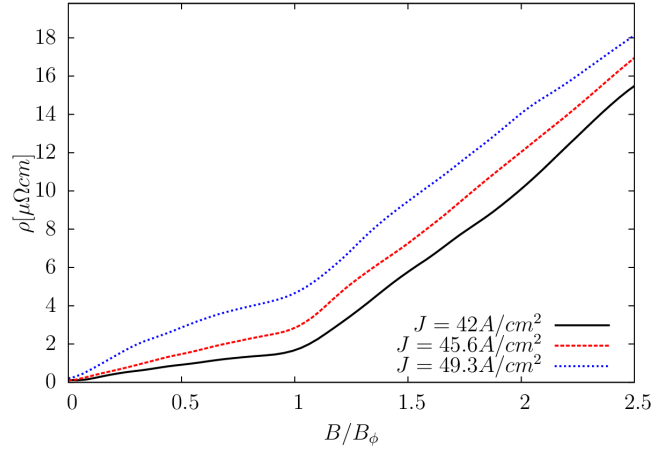}
\caption{Resistivity vs magnetic field in a Nb film with a square lattice of protruding cylinders at $T=7.65K$ (sample II). The lattice parameter is $a=650nm$ ($B_\phi=48\, G$) and the diameter and height of the cylinders is $d=340nm$ and $h_{cyl}=40nm$, respectively. The resistivity presents an approximately piecewise linear behavior with a slope changes only at the first matching field. As the current driving the vortices is increased, the slope of the low field segment increases, while it remains approximately constant for $B>B_\phi$. 
}
\label{fig:SlopesII} 
\end{center}
\end{figure}

Upon increasing the external current (i.e. the vortex driving force) or the temperature, 
the commensurability effects at $B=B_\phi/2$, $B=3B_\phi/2$ and at $B=2B_\phi$ become less pronounced and disappear (see figures \ref{fig:SlopesT} and \ref{fig:SlopesII} ). In the parameter regime where the commensurability effect at $2 B_{\phi}$ is not observed, the slope of $\rho(B)$  increases with $J$ and $T$ for $B<B_\phi$ but remains approximately constant for $B>B_\phi$. The numerical simulations show that the average single vortex velocity $\langle v\rangle_{sv}$ increases much faster than the velocity of the kinks with increasing $J$, giving rise to a different rate of increase in the slope of the magnetoresistance for magnetic fields below or above the matching field $B_\phi$.

\section{Model and Methods} \label{sec:methods}
Molecular dynamics has proven to be a very successful technique to describe the static and dynamic properties of vortex systems in a variety of experimental situations~\cite{PhysRevB.58.6534, PhysRevB.80.140514}. Here we model the magnetoresistance experiments described in the previous section. We consider an effective two dimensional model to describe the vortices in the superconducting film. The overdamped equation of motion for the $i$-th vortex is given by 
\beq
\eta \vec{v}_{i} = \vec{f}_{i}^{vv}  - \vec{\nabla} U\big(\vec{r}_{i}\big) + \vec{F}_{C}. 
\eeq
where $\vec{v}_{i}=\dot{\vec{r}}_i=(\dot{x}_i,\dot{y}_i)$ is the velocity, $\eta$ is the Bardeen-Stephen friction,
\beq
\vec{f}_{i}^{vv}=\sum_{j\neq i} \vec{f}_{ij}, 
\eeq
is the force due to the interaction with the other vortices, $U\big(\vec{r}_{i}\big)$ is the external potential, and $\vec{F}_{C}=\frac{\phi_0}{c}\vec{J}\times \hat{z}$, a driving force generated by the transport density supercurrent $\vec{J}$.

In the limit $\lambda \gg \xi,$ we have
\beq \label{rango}
\vec{f}_{ij} = f_{0}K_{1}\left(\frac{r_{ij}}{\lambda}\right) \frac{\vec{r}_{ij}}{r_{ij}}.
\eeq
where $\vec{r}_{ij} = \vec{r}_{i}-\vec{r}_{j}$, $K_{1}(r/\lambda)$ is the modified Bessel function of the first kind, and $f_{0} = \frac{\phi_{0}^2}{8\pi^{2}\lambda^{3}}$. Since $K_{1}(r/\lambda)\propto \exp(-r/\lambda)$ for $r \gg \lambda $, we cut off the vortex-vortex interaction for distances larger than $6\lambda$. 
In what follows we take $\lambda$, $\frac{\lambda \eta}{f_{0}}$ and $f_{0}$ as distance, time and force units, respectively. 

We model the interaction between the $i$-th vortex and the square lattice of protruding cylinders with the potential: 
\beq \label{eq:potcyl}
U(\vec{r}_{i}) = \frac{a\,F_p}{2\pi} \left[\cos\left(\frac{2\pi x_i}{a}\right)+\cos\left(\frac{2\pi y_i}{a}\right)\right] ,
\eeq
where $a$ is the lattice parameter of the cylinders structure, and $F_P$ is a function of $a$ and $\lambda$. \footnote{Following Berdiyorov {\it et al.}~ \cite{PhysRevB.77.024526} we expect the interaction with the cylinders to behave as $\sim f_0 e^{-r/\lambda}$ as a function of $\lambda$ and the vortex-cylinder distance $r$. Using this form, the potential along paths parallel to the rows of cylinders is approximately described by a cosine function. We assume that in the range of temperatures, currents and fields studied the vortices do not penetrate the cylinders, so that the potential considered in Eq. (\ref{eq:potcyl}) describes the interstitial region where the vortices move. }

To simulate the magnetoresistance measurements we consider a system with periodic boundary conditions and an external force along one of the principal axes of the square lattice (the $\hat{y}$ direction).
We start with $n_0$ vortices, add at random positions $\Delta n$ vortices and perform $4000$ molecular dynamics (MD) time-steps ($\Delta t=0.007$) to reach a stationary state. We average the quantities of interest over $6000$ additional MD steps. In particular we calculate the {\it total average velocity} $\langle \vec{V}\rangle=\sum_i^{n_V} \langle\vec{v}_i\rangle$ whose projection $\langle V_y\rangle$ along the y-axis is proportional to the measured voltage drop across the sample associated with the motion of the vortices. We perform the simulations in a $30 \times 30$ square array of cylinders and use $n_0 = \Delta n = 22$. The curves $\langle V_y\rangle$ vs $n_V/n_P$ presented are the result of averaging over ten realizations.

In the low vortex density limit the behavior of $\langle V_y\rangle$ can be understood assuming that the inter-vortex interaction is negligible. We may then calculate the average velocity as $\langle V_y\rangle_{NI} = n_V\langle v \rangle_{sv}$ where $\langle v \rangle_{sv}$ is the mean velocity for an isolated vortex in the system, and determines the slope of  $\langle V_y\rangle$. A single vortex in the potential generated by the cylinders follows a rectilinear path parallel to the $y$ axis equidistant to two rows of cylinders. Along these paths, the pinning potential has the form
\beq \label{eq:1Dpot}
U_{1D}(y)=  \frac{a\,F_P}{2\pi}\cos(2\pi y/a ).
\eeq
with an associated pinning force $F_{1D}(y)=-\partial U_{1D}(y)/\partial y$.
For  $F_C$ larger than the maximum pinning force $F_P = \max[F_{1D}(y)]$, the average velocity can be calculated integrating the equation of motion: 
\begin{equation}\label{fza}
\langle v \rangle_{sv} = \frac{a}{\int_0^a \frac{dy}{[1-F_{1D}(y)/F_C]}}F_C,
\end{equation}
which can be evaluated for the potential of Eq. (\ref{eq:1Dpot}) to obtain
\beq \label{eq:vsv}
\langle v\rangle_{sv} = \sqrt{F_C^2-F_P^2}.
\eeq
Equation  (\ref{eq:1Dpot}) shows that $\langle v\rangle_{sv}$ vanishes as $\sqrt{F_C-F_P}$ for $F_C- F_P \to 0$, which is a generic behavior for potentials having a quadratic maximum in the pinning force.

\section{Two-dimensional simulations} \label{sec:simulations}
\begin{figure}[t]
\begin{center}
 \includegraphics[width=9 cm]{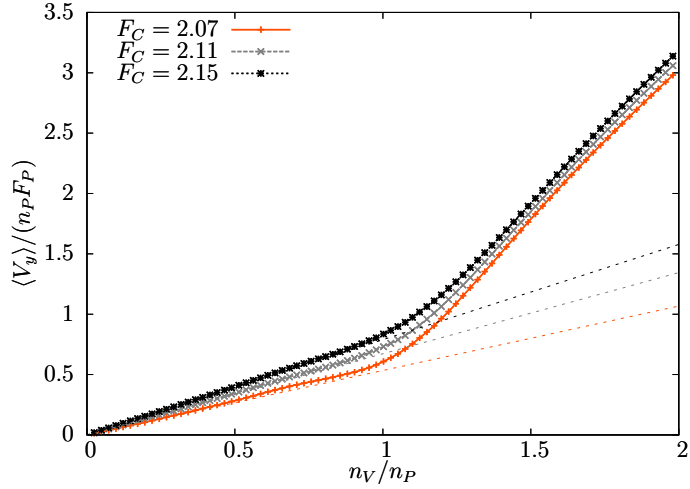}
\caption{Total average vortex velocity vs the number of vortices in a $30 \times 30$ square array of cylinders with lattice parameter $a = 2$. For Lorentz forces $F_C$ larger than $F_{P}$ and $n_V/n_P < 1$ the average velocity is well described by the single vortex value. At $n_V = n_P$ an increase in the slope is observed which is related to the propagation of kinks for $n_V>n_P$. The  dotted style lines present the results expected for a single vortex (see text).} 
\label{vel}
\end{center}
\end{figure}

Using the procedure described in the previous section, we performed numerical simulations of the magnetoresistance measurements calculating $\langle V_y\rangle$, which is proportional to the dissipation, as a function of the number of vortices $n_V$, which is proportional to the external magnetic field. The main result (see figure \ref{vel}) is the observation of a piecewise linear increase in $\langle V_y\rangle$ as a function of the number of vortices $n_V$. This behavior is obtained for a wide range of values of the penetration length, including the estimated value for the samples studied in section \ref{sec:mr} ($\lambda \sim 350 nm$) \footnote{The value of the penetration length for Nb at low temperatures~\cite{lambdaT0,PhysRevB.75.064508} is $\lambda(T\to 0)\sim 44nm$. For sample I, using the two fluid approximation we obtain $\lambda = (350 \pm 120) nm$ at $T/T_c = 0.996$.},
and reproduces qualitatively our experimental results.

In figure \ref{vel} we present $\langle V_{y}\rangle$ as a function of $n_V/n_P$ for different values of the external force $F_C$. For forces over a threshold $F_C > F_P$ and low vortex densities ($n_V\ll n_P$), the average velocity increases linearly with the number of vortices in the system following the expected behavior for a single vortex (thin lines in figure \ref{vel}). 
An approximately single vortex behavior is observed in $\langle V_y\rangle$ up to the first matching field $n_V=n_P$ where there is a change in the slope of the curves. 
For $n_V>n_P$, $\langle V_{y}\rangle$ presents again an approximately linear behavior but with a higher slope. The deviation of $\langle V_y\rangle$ from the single vortex result for $n_V>n_P$ indicates a more active role of the vortex-vortex interactions in this regime and is associated to the presence of moving discommensurations (kinks).

In qualitative agreement with the experimental observations (see figure \ref{fig:SlopesII}), at low vortex densities there is a large relative increase in the slope of the curves for a moderate increase in $F_C$, while the slope remains relatively constant for $n_V>n_P$. The large relative increase in $\langle v\rangle_{sv}$ with increasing $F_C$ can be easily understood from Eq. (\ref{eq:vsv}) in the regime $\Delta_F=F_C-F_P\ll F_C$. 

\begin{figure}[t]
\begin{center}
 \includegraphics[width=9 cm]{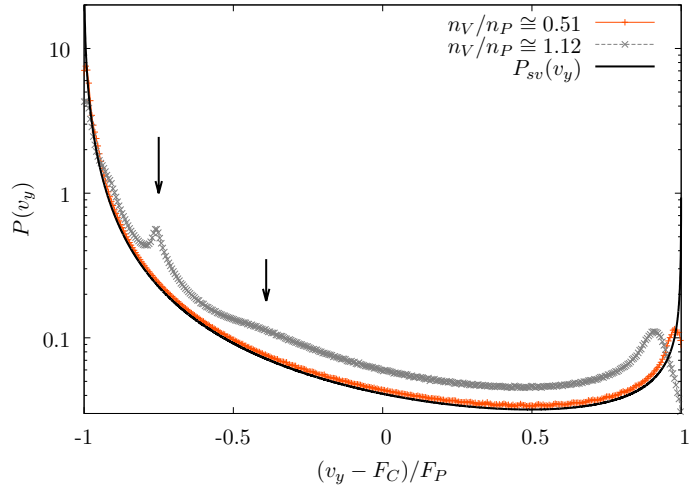}
\caption{Vortex velocity distribution $P(v_y)$ vs $(v_y-F_C)/F_P$ for $F_C = 2.07$ and $F_P = 2$ and $a = 2$.
For densities $n_V/n_P < 1$, $P(v_y) \cong P_{sv}(v)$, while for $n_V/n_P > 1$ there are two additional peaks (see arrows) and the probability of finding a vortex having a velocity $F_C - F_P$ decreases.}
\label{fig:dist}
\end{center}
\end{figure}

Figure \ref{fig:dist} presents the vortex velocity probability distribution $P(v_y)$ for the lowest value of the force considered in figure \ref{vel}. For low vortex densities $n_V\ll n_P$, $P(v_y)$ is well described by the single vortex result $P_{sv}(v)$ (see \ref{app:sine}), signaling a weak effect of vortex-vortex interactions, in agreement with the results for $\langle V_y\rangle$. The distribution is highly peaked at two characteristic velocities: a `slow' velocity $F_C-F_P$ and a `fast' velocity $F_C+F_P$. %For the parameters in Fig. \ref{dist} this is the case for values as large as $n_V = 0.54~ n_P$. 
For vortex densities above the first matching field $n_V>n_P$, a clear deviation from the single vortex result is obtained as $P(v_y)$ develops two new peaks. As anticipated in section \ref{sec:mr}, for $n_V\gtrsim n_P$ each additional vortex generates a kink that travels faster along the sample than the individual vortices. The two additional peaks in $P(v_y)$ are associated to the typical velocities of a vortex when it is participating in the kink's movement.   

\begin{figure}[t]
\begin{center}
 \includegraphics[width=9 cm]{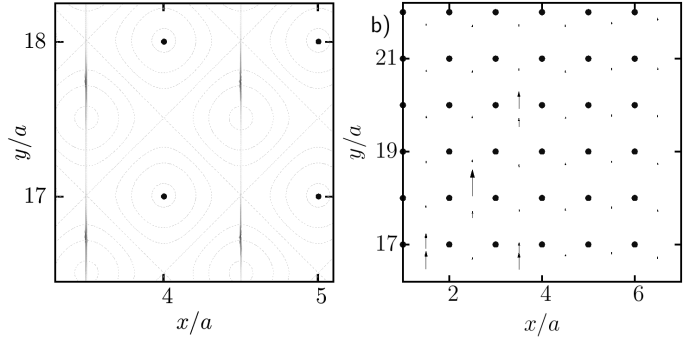}
\caption{
a) Time averaged density map of the vortex positions, the darker shades of gray indicate a higher probability of finding a vortex at a given position. The dotted style lines are equipotential contour lines of the pinning potential (\ref{eq:potcyl}) and the dots indicate the potential maxima. The density is calculated averaging over 10000 simulation time steps for $F_C=2.07$, $F_P=2$, $a=2$, and $n_V=1.2 n_C$ in a $30\times 30$ simulation box. The vortices move along approximately rectilinear paths between energy maxima and spend most of the time climbing the potential barrier.
b) Snapshot of the vortex configuration where the arrows indicate the vortex velocities. The vortices having a larger velocity are associated to the presence of moving discommensurations.}
\label{fig:paths}
\end{center}
\end{figure}

In the range of magnetic fields studied and for values of the penetration length where a piecewise linear behavior is observed in the magnetoresistance, an analysis of the vortex trajectories shows that they follow approximately rectilinear paths along channels defined by pairs of cylinder rows. This is illustrated in figure \ref{fig:paths}(a) where a time average of the vortex positions is plotted for a regime with $n_V>n_P$. 
Figure \ref{fig:paths}(b) shows a typical configuration of the vortex velocities at a given time. Most vortices move with a `slow' velocity, as expected from $P(v_y)$, while a few move with a faster velocity. These faster moving vortices are associated to the presence of kinks. As we will show in the following section, the observed piecewise linear behavior of the magnetoresistance can be understood within a purely one-dimensional model.

\section{One-dimensional toy model} \label{sec:toymodel}
In the two-dimensional numerical simulations presented in the previous section the vortices moved along approximately rectilinear paths. 
In this section we consider a one-dimensional toy model that provides a simple qualitative picture of the kink dynamics. 
For simplicity we consider a periodic triangular potential such that the force on a vortex is given by:
\[
F(y)=-F_P\,\,sign[\cos(2\pi y/a)].
\]

At zero temperature for vortices driven by an external force $F_C>F_P$, the single vortex average velocity is:
\beq \label{eq:avg1v}
\langle v\rangle_{sv} =F_C-\frac{F_P^2}{F_C}=\frac{(F_C-F_P)(F_C+F_P)}{F_C},
\eeq
and the velocity probability distribution results:
\beq \label{eq:dist}
P(v)=\frac{F_C+F_P}{2 F_C}\delta(F_P-F_C)+\frac{F_C-F_P}{2 F_C}\delta(F_P+F_C).
\eeq
The presence of two characteristic velocities for the motion of an isolated vortex: a `slow' velocity $F_C-F_P$ and a `fast' velocity $F_C+F_P$ is the main ingredient for the observation of the kinks. 

We also use a simpler form of vortex-vortex interaction potential including a cut-off:
\beq \label{eq:potential}
v_{int}(u) = \lambda f_0\,\left(-\log|u|-\frac{u^4}{4} + u^2-\frac{3}{4}\right),
\eeq
where $u=\frac{2}{5}r_{12}/\lambda$, and the cut-off is at $u_c=1$. The precise form of the interaction is not important for a qualitative description of the magnetoresistance provided that vortex-vortex interactions are negligible at large distances.

At the first matching field the vortices are evenly spaced and synchronized, and the probability distribution of the velocity is given by the single vortex result of Eq. (\ref{eq:dist}).
For $\Delta_F=F_C-F_P\ll F_C$ there is a large probability of finding the vortices with a slow velocity $\Delta_ F$. \footnote{For a sinusoidal potential the $\delta$-functions are replaced by square root divergences but the physical picture remains the same (see \ref{app:sine}).}
This is because in the process of going over a period of the potential, the vortices spends most of the time traveling at a slow velocity and it is the slow velocity which dominates their average velocity. 

\begin{figure}[t]
\begin{center}
\includegraphics[width=8 cm,angle=0,keepaspectratio=true]{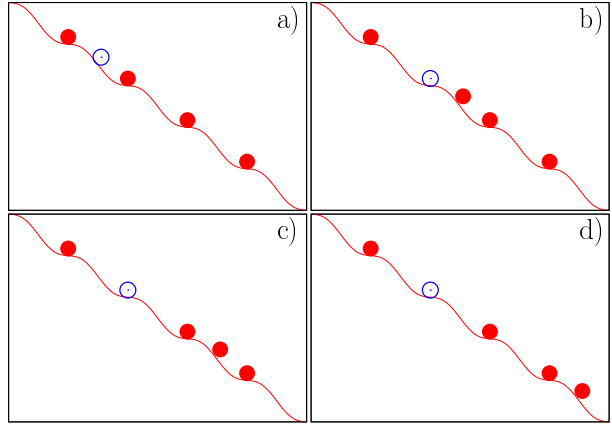}
\caption{Schematic representation of a moving discommensuration in a 1D system. The temporal sequence is starts at a) and ends at d). Initially there are as many vortices (filled red dots) as periods in the total potential (solid line) which is given by the pinning potential and the driving potential. A single vortex added to the system (open blue dot) generates a cascade of vortex displacements.}
\label{fig:soliton}
\end{center}
\end{figure}

In this situation, adding a single vortex to the system generates a defect that produces a cascade of displacements (see figure \ref{fig:soliton}). The additional vortex pushes its first neighbor which is in a slow velocity region increasing its velocity. The extra vortex stays trapped in a region with a slow velocity while its neighbor now with a fast velocity approaches the next vortex in the chain to help it in turn go faster along the slow velocity region. 
 This defect propagates faster than the average velocity of the vortices and is equivalent, in its contribution to the magnetoresistance, to the addition of a vortex with a speed $v_k>\langle v\rangle_{sv}$.

Increasing further the number of vortices increases the number of kinks and the dissipation. The increase is however larger than the one obtained for $B<B_\phi$ as the slope of $\rho$ is given by the kink velocity $v_{k}$.
The kinks repel each other and tend to be evenly spaced but this interaction does not lead to a significant change in their average velocity.  
While the slope of the first segment ($B<B_\phi$) is dominated by the single vortex result for the average velocity $\langle v\rangle_{sv} \sim F_C-F_P$, the slope for fields above the first matching field is given by the speed of the kink which is typically a fraction of $F_C$. For $F_C\sim F_P$, a small change in $F_C$ can lead to a large relative change in the slope of the first segment while producing a small relative change in the slope of the second, as is observed experimentally (see figure \ref{fig:SlopesII}). 

Increasing further the number of vortices up to the second matching field, assuming that the 1D model still gives a qualitative picture for these vortex densities \footnote{This is possible, provided the range of the vortex-vortex interactions, given by $\lambda$ is short enough.}, can lead to a second kink in the magnetoresistance. 
\begin{figure}[t]
\begin{center}
 \includegraphics[width=9 cm]{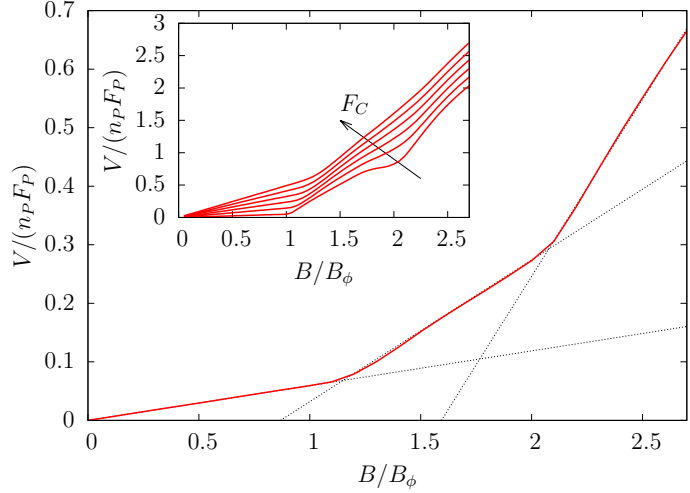}
\caption{Total velocity for vortices moving in a triangular potential. Two changes in the slope of the velocity are observed as a function of the number of vortices. The dotted style lines are linear fits and serve as guides to the eye. The parameters are $F_C=1.03\,F_P$, $a=4$, $F_P=10$. Increasing $F_C$ the commensurability effect at $B=2B_\phi$ becomes weaker and eventually disappears. This is shown in the inset for a different set of parameters $F_C=1.025\,F_P,1.075\,F_P,\ldots,1.275\,F_P$, $a=4$, $F_P=1.33$.}
\label{fig:1D2kinks}
\end{center}
\end{figure}
This can be observed in figure \ref{fig:1D2kinks} where there are two changes in the slope of the total velocity. The dynamics above the second matching field is similar to the one above the first matching field. At the second matching field the vortices form a commensurate structure and an additional vortex generates a moving defect.   

Decreasing the pinning force $F_P$ or increasing the Lorentz force $F_C$ reduces the time spent by the vortices in the slow velocity regions and eventually destroy the possibility of having a second slope change  at $B=B_\phi$ in the magnetoresistance. This behavior is shown in the inset of figure \ref{fig:1D2kinks} where we present simulations of the magnetoresistance for different values of $F_C$.

\section{Summary and Conclusions} \label{sec:concl}
We measured the electronic transport through superconducting Nb films having a periodically modulated thickness as a function of the magnetic field. We observed a piecewise linear behavior of the resistivity that we interpreted in terms of different dynamic regimes of the superconducting vortices. The thickness modulation of the film generates a periodic potential for the vortices which induces commensurability effects in the vortex flow. Numerical results for the vortex dynamics indicate the presence of different moving discommensuration (or kink) regimes. In the regime of parameters where there is a good qualitative agreement with the experimental results, the kinks observed are well described considering independent one-dimensional vortex chains formed between pairs of neighboring rows of cylinders. 

The generation of a thickness modulation using lithographic techniques offers a unique opportunity to control precisely the intensity, geometry and topology of the pinning potential. Ginzburg-Landau calculations allow in principle for a calculation of the pinning potential surface, making these systems a fertile ground for experiment-theory comparisons and the study of different dynamical regimes and transitions.

\ack
We thank Francisco de la Cruz, Hern\'an Pastoriza and S. Bustingorry for illuminating discussions at the early stages of this work. We thank Juan Z\'arate, Ignacio Artola and Cesar Chialvo for assistance and advice in the sample fabrication process.
We acknowledge financial support from PIP 2009-1821, PICT 2007-819 and PICT 2007-824 of CONICET, PICT-Bicentenario 2010-1060 of the ANPCyT, and grants from Universidad Nacional de Cuyo.
\appendix
\section{1D cosine potential} \label{app:sine}
For a cosine one dimensional (1D) potential
\beq
U(y)=  \frac{a\,F_P}{2\pi}\cos(2\pi y/a ),
\eeq
the equation of motion of a single vortex
\beq
\dot{y}=F_C-F_P\sin(2\pi y/a)
\eeq
 can be readily integrated to obtain:
\beq
y(t)=\frac{a}{\pi}\tan^{-1}\left[\frac{F_P + \langle v\rangle_{sv}}{F_C}\tan\left( \frac{\pi \langle v\rangle_{sv} t}{a} - \frac{\pi}{2}\right)\right]
\eeq
valid for $-a/2<y(t)<a/2$, where  $\langle v\rangle_{sv}=\sqrt{F_C^2-F_P^2}$ is the average velocity of a vortex over a period of the pinning potential. The probability distribution of the vortex coordinate $y$ over a period can be calculated using the relation $P(y)=\frac{P(t)}{\dot{y}(t)}$
where $P(t)=\frac{\langle v\rangle_{vs}}{a}$:
\beq
P(y)=\frac{1}{a}\frac{\langle v\rangle_{vs}}{F_C-F_P\sin(2\pi y/a)}.
\eeq
Using $P(\dot{y})=\frac{P(y)}{\ddot{y}(t)}$ (here two branches must be added because the velocity is not a monotonous function of the coordinate) we obtain the single vortex velocity probability distribution
\beq \label{eq:svpv}
P_{sv}(v)=\frac{1}{\pi v} \sqrt{\frac{(F_C-F_P)(F_C+F_P)}{(v-F_C+F_P)(F_C+F_P-v)}},
\eeq
which has square root divergences at $v=F_C\pm F_P$. As in the toy model described in section \ref{sec:toymodel}, for $0<F_C-F_P\ll F_C$ there is a large probability of finding the vortex with a slow velocity $v\sim F_C-F_P$. This is a central ingredient for the physics of moving discommensurations described in the main text.

\bibliography{vortex}{}

\end{document}